# Cloud-Based Big Data Management and Analytics for Scholarly Resources: Current Trends, Challenges and Scope for Future Research

Samiya Khan, Kashish A. Shakil, and Mansaf Alam

**Abstract**—With the shifting focus of organizations and governments towards digitization of academic and technical documents, there has been an increasing need to use this reserve of scholarly documents for developing applications that can facilitate and aid in better management of research. In addition to this, the evolving nature of research problems has made them essentially interdisciplinary. As a result, there is a growing need for scholarly applications like collaborator discovery, expert finding and research recommendation systems. This research paper reviews the current trends and identifies the challenges existing in the architecture, services and applications of big scholarly data platform with a specific focus on directions for future research.

——————————— ◆ ———————————

## 1 INTRODUCTION

THE digital world is facing the aftermath of data explosion, which has led to the coining of terms like data deluge. In simple terms, data deluge is a phrase used to describe the excessively huge volume of data generated at a regularly increasing basis in the world. Organizations are overwhelmed by the processing and storage requirements of such large volumes of data. With that said, another implication of the data deluge is that it has made the scientific method completely obsolete.

Traditionally, the scientific method for solving a problem requires definition of the problem, proposal of a solution and collection of data that can solve or support a solution to the problem. However, there is abundant, easily accessible data, present today. In order to make use of this reservoir of data, researchers need to ask the right questions that this data can answer for them. Therefore, the approach is changed from 'ask the question; collect data' to 'frame a question that the available data can answer'. In order to support this new approach, particularly for scholarly resources, big scholarly data analytics has come into existence.

Scholarly documents are generated on a daily basis in the form of research documents, project proposals, technical reports and academic papers, in addition to several other types of documents, by researchers and students from all over the world. Moreover, there have been several initiatives by Governments and Organizations to digitize existing academic resources [7][8][9]. It is this huge reservoir of academia data that is popularly referred to as 'scholarly data'. However, it is important to note that this is a generalized description and the definition may vary from one scholarly community to another. For instance, Google Scholar does not count patents as a scholarly resource.

With that said, the abundance of data sources makes large-scale analysis of scholarly data possible and feasible. However, commercially available solutions in this area are rather limited. There have been several research efforts in the field of academic search engines. Some of the popular search engines include CiteSeerX [1] and Google Scholar [2]. In addition, assessment and benchmarking tools like Microsoft Academic Search [3] and AMiner [4] also exist. While these are primary sources of scholarly data, BASE [5] or Q-Sensei Scholar [6] are services that depend on secondary sources of preprocessed data.

Big Scholarly Data Analytics have far-reaching implications on the ease with which research is performed. Primarily, analytics for big scholarly data can be divided into four categories namely, research management, collaborator discovery, expert finder systems and recommender systems. Such analytics have gained immense importance and relevance lately particularly with the advent of multi-disciplinary research projects.

Such projects have increased the scale and complexity of research problems manifold and emphasize on the pressing need for collaboration among researchers as well as institutes or organizations. Research collaboration is not a neo-concept. However, there has been a recent shift in the manner in which collaborations are initiated. Traditionally, researchers and scholars used to meet periodically in conferences and symposiums to explore new research domains and possibility for collaborations.

————————————————

• *Samiya Khan is with the Department of Computer Science, Jamia Millia Islamia, New Delhi, India. E-mail: samiyashaukat@yahoo.com.*
• *Kashish A. Shakil is with the Department of Computer Science, Jamia Millia Islamia, New Delhi, India. E-mail: shakilkashish@yahoo.co.in.*
• *Mansaf Alam is with the Department of Computer Science, Jamia Millia Islamia, New Delhi, India. E-mail: malam2@jmi.ac.in.*

With the increasing popularity of Internet, these platforms have been complemented with academic search-oriented web engines like Google Scholar and academic social networking portals like ResearchGate [35] and Academia [36]. While these platforms allow researchers to follow each other's research activities and interests, they have also created a sense of realization in the research community that the final published article is merely a milestone in research.

Other aspects of research like dataset used and supporting material considered for the research are equally important. This is one of the reasons for the staggering rise of interest in research data management. Although, research management, collaborator discovery and expert finding remain popular analytics applications, several other useful applications can be implemented to make optimal use of the heaps of scholarly data available to provide personal, local and global insights in the research work performed in this area.

This research paper aims to study the current trends in cloud-based data management and analytics of big scholarly data and identify the challenges that continue to exist in the different phases of the system. Besides this, it shall also give an analysis of the scope for future research in this field. The rest of the paper has been organized in the following manner: Section 2 gives an introduction to cloud-based big data analytics and reviews existing platform for big scholarly data, which also serves as the base for future research work in big scholarly data analytics.

The trends, challenges and research directions have been classified under three main categories namely, data management, analytics and visualization. Section 3, Section 4 and Section 5 cover these three categories in detail. The challenges discussed in the three sections mentioned above constitute only technical challenges. This field of study also suffers from some non-technical challenges, which have been described in a Section 6. The paper concludes with a remark on the scope of research in this area and future research directions.

## 2 BACKGROUND AND METHODOLOGY

Big data analytics is a vast field that has found applications in diverse domains and studies. Some of the most impactful researches that have merged big data analytics with other fields of study include business analytics, multi-scale climate data analytics [11], banking customer analytics [14], smart cities [16], recommender systems for ecommerce [13], social media analytics [12], healthcare data analytics [15], intelligent transport management systems [18] and railway assets management system [17].

Evidently, the type of data analytics required for fulfillment of the needs of specific fields is different. Chen and Zhang [19] provided an extensive survey on the tools, techniques and technologies used for big data analytics. The commonest mathematical tools used for analysis of data include fundamental mathematical concepts, statistical tools and methods for solving optimization problems. On the other hand, analytical techniques required for making big data analytics feasible and usable for the end users include machine learning, data mining, signal processing, neural networks and visualization methods.

In order to implement the techniques mentioned above, MapReduce and Hadoop [20] has been identified as the most effective and efficient framework. Hadoop is an open-source implementation of the MapReduce programming model that allows distributed processing of a huge volume of heterogeneous data using commodity machines. Although, the research work paid little heed to deploying Hadoop on the Cloud, it has indicated that Cloud Computing is one of the proposed technologies for backing big data analytics applications.

Cloud computing promises to be a good solution to the big data problem considering the scalability and elasticity that it offers [25]. However, the viability of this synergistic model is yet to be explored and tested. Big data computing, particularly in the cloud environment, itself suffers from some inherent challenges [24][26].

Assuncao et al. [21] presented the technical and non-technical challenges associated with cloud-based big data analytics, with specific emphasis on the relevant work that has been performed in each sub-area. While the latter deals with issues concerning the management and adoption of these solutions, the former has been further classified into three categories namely, data management, model building and scoring and visualization and user interaction. A typical workflow for big data analytics given by [21] has been illustrated in Fig. 1.

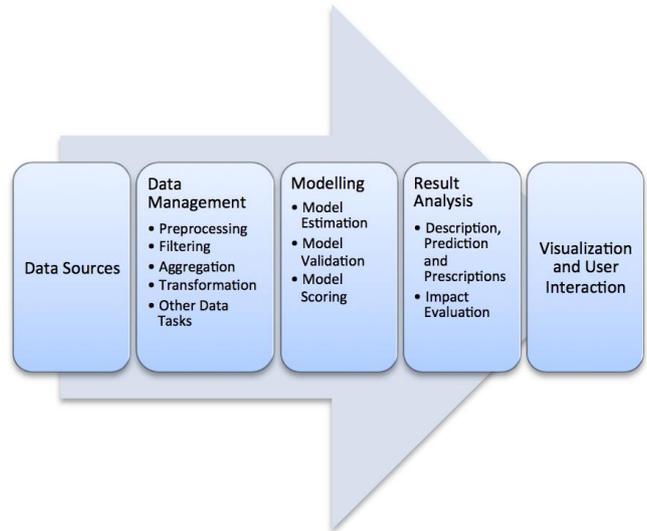

Fig. 1. Workflow for Big Data Analytics

One of the pioneering research projects in the field of Big Scholarly Data is CiteSeerX. Wu et al. [10] presented the platform for big scholarly data, which proposes to move the then-existing system of CiteSeer to a private cloud. Teregowda and Giles [160] elaborated on this in a detailed report on scaling SeerSuite in the cloud environment. The platform is divided into three components namely, architecture, services and applications. The system makes use of Crawl Cluster, HDFS, NoSQL and MapReduce for implementation.

The proposed system can broadly be divided on the basis of user interaction into two sections – frontend and backend. The frontend includes load balancers and web servers. This interface allows user to interact with the system, takes their requests and communicates the results back to the users. On the other hand, the backend performs crawling of web sources for relevant data, extraction of information from raw data and ingestion of information into the system to support applications like research management, collaborator discovery and expert finding, in addition to several others. An illustration of the big scholarly data platform, proposed by Wu et al. [10], has been presented in Fig. 2.

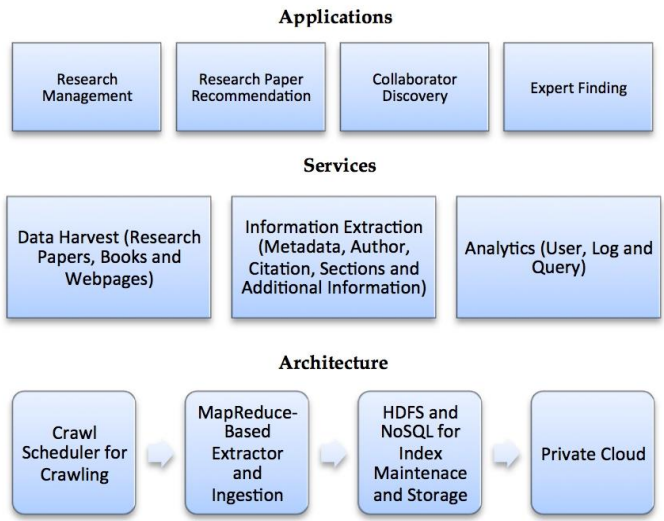

Fig. 2. Big Scholarly Data Platform

On the basis of the architecture, challenges and research directions proposed by Wu et al. [10], this research divides the challenges presented by cloud-based analytics of big scholarly data into technical and non-technical challenges. Research papers under each category have been analyzed using the qualitative research methodology to provide an extensive survey on cloud-based big scholarly data platform. The technical challenges are further divided on the basis of the functionality to which the challenges belong. The three categories include data management, analytics and visualization, which have been covered in the sections that follow.

## 3 SCHOLARLY DATA MANAGEMENT

Data is generated in many diverse forms in any scholarly platform. One of the primary sources of data is the huge reservoir of existing scholarly documents on the Internet. In addition to this, there are author webpages, academic social networks and secondary sources of scholarly information like institution and organization webpages that also render significant data for a comprehensive analysis of the scholarly community. Evidently, there are several sources of data, providing different types of information. Moreover, this data is continuously updated, appended and removed. Challenges in data management can be further divided into four sub-categories: (i) big data characteristics (ii) data acquisition and integration (iii) information extraction (iii) data preprocessing (iv) data processing and resource management. The different facets of data management of big scholarly data have been discussed below.

### 3.1 Big Scholarly Data Characteristics

Big data is traditionally characterized by three main features namely volume, variety and velocity. It can be derived from the meaning of these words that volume characterizes the size of data, variety symbolizes the types of data included and velocity indicates the rate of data generation.

The volume of data can be assessed by evaluating the size of scholarly documents available on the web as raw data. Khabsa and Giles [23] estimated that the number of English scholarly documents available on the Internet is approximately 114 million and this value is incremented at a daily rate of tens of thousands. It is crucial mention here that this is the lower bound value. It has also been stated that the Google Scholar accommodates 87% of the total [23]. Therefore, the number of English scholarly documents on Google Scholar is around 100 million [23].

It is important to understand that the big scholarly dataset is not just limited to scholarly documents. Information extracted from raw data and linked to create citation and knowledge graphs are also significant contributors to the size, variety and volume of big scholarly data. Caragea et al. [32] gave an estimate of the big scholarly dataset maintained at CiteSeerX until May 2013. The total number of documents in the aforementioned system was approximately 2.35 million. However, this count includes duplicates and upon removal of the same, the approximate count is reduced to 1.9 million documents. In addition to this, the number of unique authors in the database is 2.4 million while the number of citations, which includes repetitions, is about 52 million.

From data size perspective, Caragea et al. [32] estimated the size of CiteSeerX to be 6TB, which is growing at a daily rate of 10-20GB. From the numbers stated above, it can be implied that scholarly dataset is indeed 'big'. Specifically, there are three main reasons why scholarly data is called big scholarly data, which are as follows:

1. Firstly, the storage and computing resources requirements of this data are too high to be provisioned by traditional architectures. For instance, common scholarly applications like collaborator discovery require services like author profiling and disambiguation. This is a computing intensive task, which requires the system to work on 'big' data. Moreover, one of the fundamental requirements of this system is smart resource allocation and scheduling.

2. Secondly, the data throughput requirements of the system need a better data processing framework and tools. The single pipeline system is the bottleneck, particularly in the case of data ingestion.

3.  Lastly, static crawling techniques do not provide the coverage and data filtering accuracy that such systems and applications require. Besides this, existing document classifier systems perform basic classification, separating academic documents from non-academic documents. For advanced applications, more sophisticated classification, on the basis of document type and subject, is required.

In addition to the standard 3V characteristics, Wu et al. [22] gave many new attributes, transforming the 3V model into the multi-V model. Additional characteristics include veracity, value, variability, validity, visibility and verdict. A Venn diagram for the multi-V has been shown in Fig. 3. The 3Vs – value, visibility and verdict – constitute the business intelligence (BI) aspects of the data concerned.

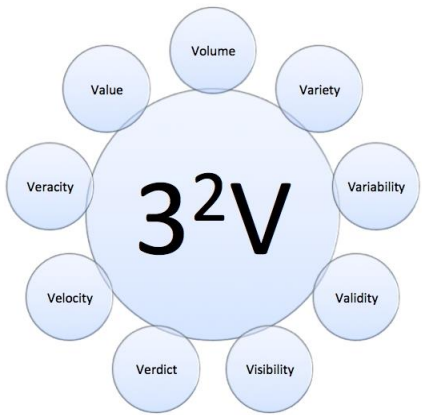

Fig. 3. Venn diagram of $3^2$V Model

The visibility characteristic provides the foresight, hindsight and insight of the data as opposed to the traditional 3Vs that only focus on insight. From the BI perspective, it is important to know if the data is capable of contributing anything substantial, which defines the 'value' of data. On the basis of analysis of the problem and its proposed solution, it is the decision makers' job to give a 'verdict'.

The statistical perspective on data is given by veracity, validity and variability. Veracity defines the trustworthiness of data while validity determines if the data has been acquired ethically and without any bias. When data complexity and variety are analyzed, the implied characteristic that comes into being is 'variability'.

It is important to note that there is limited research performed on data veracity. Data quality has a direct impact on the quality of analytics produced, which makes veracity a significant big data characteristic, particularly for critical applications [41][43]. In addition, the privacy and security aspects of cloud-based big data solutions, which are remarkably significant in view of the fact that these facets are important user concerns [46] when working in the cloud environment, are also yet to be explored in full.

Although, validity is a conceptual concept and holds little significance in the present context, variability is particularly relevant to big scholarly data. The 3Vs associated with business intelligence perspective solely depend on the ability of an organization to make use of the available data with the deployed solution. Moreover, there is no existing literature that discusses big scholarly data with respect to the statistical and business intelligence perspective.

### 3.2 Data Acquisition and Integration

The first step of the data analytics process is data acquisition, as part of which data is collected from a single source or multiple sources and integrated to form the dataset that serves as input to the analytics engine. A big scholarly dataset is an integration of many types of documents, which has been illustrated in Fig. 4. These documents are retrieved from their respective sources. The primary source of data is the web, with specialized databases like DBLP [37]. Moreover, portals like arxiv [38] and publishing houses like Elsevier [39] also provide APIs, which can be used to extract data.

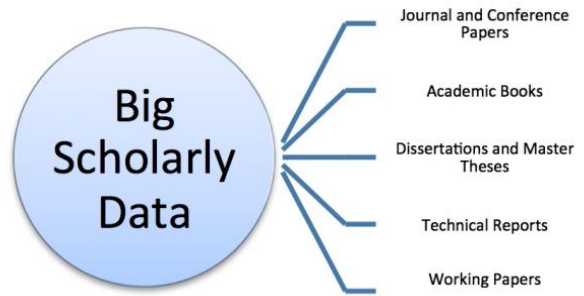

Fig. 4. Big Scholarly Dataset Composition

In order to extract data from the web, two tools can be used namely, crawling and REST APIs. CiteSeerX uses focused crawling [161], as it only requires academic documents [40]. Two crawlers, one of which performs scheduled crawling while the other crawls URLs submitted by users, which is a source of rich and dependable data, extract only PDFs. Moreover, the former satisfies the data freshness characteristic of data acquisition by keeping the database updated with latest publications.

The crawling process yields PDFs. However, the classification of documents as academic or non-academic is done as part of the document filtering process. The text of the PDF is extracted and on the basis of presence or absence of Bibliography or References at the end of the text, it is classified as academic or non-academic. Only academic documents are kept and the rest of the documents are discarded.

One of the most important facets of data acquisition is to determine if a single source is enough to get all the data required for providing accurate analysis. In order to address this concern, there have been several efforts to estimate the total size of big scholarly data, the value of which is then compared with individual statistics pub-lished by search engine and databases' owners to determine if there is a single scholarly reservoir that can serve the data needs of an analytical engine.

Several databases and Academic Search Engines exist. These systems track online scholarly documents and in the process facilitate research. There have been individual efforts to estimate the number of scholarly documents available on each of these systems, some of which have been given as system statistics by the owners of the databases and engines.

As of May 2016, the size of Web of Science was estimated to have 61 million documents [27]. However, Microsoft Academic Search (MAS) Engine was estimated to consist of more than 80 million documents [28]. PubMed and CiteSeerX are comparatively smaller data repositories and most of the documents indexed in them are also present in Google Scholar and MAS. Out of all the available sources of data, Google Scholar is considered the largest.

There have been several research efforts to determine the size of Google Scholar [68][23][70]. However, it has been established that calculating the size of Google Scholar is not the same as calculating the size of the Web [69]. Some of the researches in this area determine the citations overlap to calculate the size [29][30]. One of the latest works in this area estimate that the size of Google Scholar within the period of 1700 to 2013 is 170-175 million unique records [67].

Estimate on scholarly documents published yearly is also available. In the year 2006, 1.35 million documents were published [31] while 1.8 million documents were published in 2011 [70]. Although, Google Scholar is the largest database, the disparities in these values indicate that a single source will not be enough to create a comprehensive scholarly dataset for analysis.

A significant issue faced in this regard is that the same document may be available at several locations like author pages, sharing portals and publisher links. Therefore, different libraries or databases may have taken data from different sources. This makes it essential for the system to not just look at the data source, but also the data extracted from the source.

It is possible that the source provided by the author may allow access to the document to the author, but an automated web crawler may not be able to access the full documents. As a result, an automated web crawler faces this as the biggest challenge. Besides this, some data sources' API-based data extraction method is limited by the number of records and fields that can be extracted per query or per day [71].

As far as data integration is concerned, while integrating data from different sources is one aspect of the challenge, integrating data of different types (structured, unstructured and semi-structured) poses an even bigger problem. With specific reference to big scholarly data analytics, integration of open access sources of data like Wikipedia and Government data needs to be explored.

### 3.3 Information Extraction

Raw scholarly data is processed to extract useful information. This process is referred to as information extraction. This process has two-fold effect on the overall usage and usability of scholarly applications. Moreover, the quality of service provided by the scholarly platform is also dependent on this phase. Broadly, information extraction presents three challenges, which include:

- Accuracy
  The accuracy of the information extraction methods directly affects data quality and quality of analytics results. Therefore, it is critical to achieve as high accuracy as possible.
- Coverage
  The coverage of information extraction methods is determined by precision and recall. While achieving a good recall is important, extracting true structures is equally important.
- Scalability
  The previous challenges were general challenges faced by all information extraction methods regardless of the data on which they are being applied. Scalability is a challenge that is specific to big scholarly data owing to the large size of data to be processed. MapReduce [90] serves as a useful and viable programming paradigm for managing the scalability issue.

Primarily, four types of information need to be extracted from scholarly data namely metadata, author information, citations and section, in addition to additional information like concept hierarchies that can be derived from basic extracted information. These types along with the approaches and procedures used for their extraction have been discussed below.

#### 3.3.1 Metadata

Metadata is the first set of data that is extracted. This data is useful in view of the fact that it forms the basis for search and indexing. Typically, metadata includes title, abstract, authors, issue and volume of publication, venue, publisher, page numbers, publisher contact details, date of publication, ISBN and copyright. Several supervised machine learning-based metadata extraction methods are available.

Wu et al. [10] described the use of SVM-based metadata extraction (SVMHeaderParse [56]) for CiteSeerX. However, this method is known to work poorly for metadata extraction of books [72]. In order to address this issue, the use of active learning has been done [73]. Besides this, Lipinski et al. [96] compared many header parsers to conclude that GROBID [95] is the best parser. The study was conducted on arxiv dataset and can be tested for a large dataset.

Quality of data extracted can be improved by removing any disambiguation that may be present. Treeratpituk and Giles [89] proposed a method for disambiguation; the fundamentals of which can be applied to metadata as well. An important point to mention is that additional information about authors and scholarly documents need to be managed by the system for quality improvement. Provenance management fundamentals for electronic data can be applied to gain better control over data quality [42]. However, provenance management for big data poses several challenges [44], which will also have to be mitigated.

### 3.3.2 Author Information

Most scholarly applications require author information for analytics. Moreover, author information is usually the basis of search in academic search engines. While authors of the scholarly document are the information that is directly extracted from the document, there are many other facets of this information that are derived from this primary data. Firstly, it gives insights about co-authorship, which also forms the basis for creation of co-authorship graph. Besides this, scholarly documents also contain author information like affiliation and email addresses [86].

The content of the work can be used to map the research interests of the author. Many other types of information like venues where the author has published or presented work and detailed author information derived from the professional author webpage can be used to form a comprehensive scholarly author profile, which can be useful for advanced scholarly analytics like collaborator discovery and expert finding [86][87][88].

### 3.3.3 Citations

Apart from author information, the second type of information that comes directly from the extraction is citation data. Citation extraction can be performed using ParsCit [92], FLUX-CiM [93] and a CRF-based system [94]. Ororbia et al. [47] compared the three methods and concluded that the performance of ParsCit and CRF-based system is comparable. Besides this, it outperforms FLUX-CiM. ParsCit lacks the capability to tokenize strings beyond white space. Therefore, the mistakes made by this parser must be corrected using preprocessing heuristics to improve its accuracy.

### 3.3.4 Sections and Additional Information

Scholarly documents can either be books or research papers and technical reports, both of which are PDFs. However, the structural organization of these two types of documents is significantly different. Tuarob et al. [97] proposed a hybrid algorithm that can identify section boundaries, detect section headers and recognize the hierarchy of sections with good accuracy. However, the approach has not been tested for big data.

The main sections present in almost all research papers are Introduction, Literature Review or Related Literature, Methodology, Result, Discussion, Conclusion, Acknowledgements and References. Moreover, every scholarly document also contains figures, tables and subject-specific elements like algorithms. Each of these sections can be extracted to give useful insights about the research work.

Acknowledgements contain key information about key people, organizations and funding agencies involved in the project. Khabsa et al. [98] developed AckSeer, which is a search engine and repository of extracted acknowledgments sections from documents. A challenge specific to the extraction of acknowledgements section is that of entity resolution. A person, organization or company may be referred to by many name variations. As a result, one canonical name can be used to cluster several entities, giving rise to name-entity resolution problem [99].

Figures form the second most important structural component of scholarly documents. Carberry et al. [103] insisted on the fact that figures are rich sources of information. Existing work in this area is limited to figure caption and associated metadata extraction [101], metadata-based search [102] and data extraction from 2D line graphs and curves [104].

Another significant effort in this field was made in the form of VizioMetrix [116], which is a scholarly platform that processes scholarly documents so as to classify the figures present in them and use the same for advanced information retrieval and bibliometric analysis. There is scope for extensive research in this field. Firstly, the data extraction functionality specific to figures can be extended to other complex graphs and mathematical structures. Besides this, vector image extraction also remains a subject of research interest in this field.

Results are commonly tabulated for summarization in scholarly documents. This makes tables an important and rich source of data, specific to the document. TableSeer [100] is a table-based search that extracts tables and the metadata associated with the same, which is then used for providing search functionality. Computer Science research documents contain specific sections like pseudo-codes and algorithms, which play an instrumental role in mapping research growth and evolution.

In order to detect pseudocodes, Tuarob et al. [105] proposed a hybrid algorithm that makes use of a hybrid of machine learning-based and rule-based approach for detecting pseudocodes. This approach performed better than individual approaches and has been adopted in AlgorithmSeer [55], which is an algorithm search engine. AlgorithmSeer also supports simple heuristic-based linking of algorithms.

However, this research can be extended to support semantic analysis and evolution of algorithms. Besides this, these concepts can also be applied to study the impact of algorithms on one another. Lastly, the prototype implementation assumes that algorithms of the same section are linked. This assumption is yet to be statistically proven. Tuarob et al. [54] also proposed the use of algorithm co-citation network to detect algorithmic level of similarity, which can further be extended to implement algorithm recommendation engines.

The citation network will not be complete unless book citations are also considered. In fact, books form the largest part of the citation network [109]. In view of this, books can be viewed as the most significant and voluminous part of big scholarly data. Gao et al. [110] reviewed structure extraction in books and proposed that extraction of ToC and metadata from books can be seen as a matching problem on bipartite graph. A book mostly contains ISBN, which can be extracted by matching the string ISBN in the extracted text.

Wu et al. [56] gave a hybrid approach using SVM-based extractor and rule-based extractor for extracting authors and title of a book. Two sections that differentiate books from other scholarly documents is the presence of table of contents (TOC) [108] and indexes [106][107], usually present at the back of the book. Moreover, unlike research papers, books have a bibliography or references section at the end of each chapter. Therefore, the book needs to be scanned in full for references and citations.

Recent developments in the study of scholarly documents have introduced the concept of scholarly knowledge graph, which shall link all the entities and information of the scholarly ecosystem. When it comes to organizing knowledge, one of the tools that can be put to use is concept hierarchy. Wang et al. [57] presented a recent work on the extraction of concept hierarchies in books. The proposed approach captures the global coherence and local relatedness in the book by extracting concepts in each chapter and constructing concept hierarchy. Wikipedia has been used as a resource for extraction of concepts. This work can be extended to use multiple books for creation of domain-specific concept hierarchies, which can further be used in scholarly applications.

### 3.4 Clustering Documents and Linking Entities
Once information has been extracted, the next step in the process is to link data. Basically, existing data needs to be linked to this newly extracted data. However, this process includes several sub-processes, which are discussed below.

#### 3.4.1 Elimination of Duplicates
Duplicates can be exact duplicates or near-duplicates. In order to eliminate exact duplicates, the SHA1 values of newly extracted documents are matched with that of existing documents and key mapping algorithm can be used for getting rid of near duplicates [47]. The Key Mapping algorithm is also used to align papers to citations. This method is adopted to get key information like date of publications and copyright for papers directly from the citation string instead of extracting this information from the PDF.

Once near duplicates are detected, they can be placed in the same cluster. This type of clustering is called metadata-based clustering. For any documents that aren't near duplicate, a new cluster corresponding to each of this document has to be created. This type of clustering suffers from an inherent drawback. The quality of clustering depends on the accuracy of the method used for metadata extraction. William and Giles [91] explored better near duplicate detection method that makes use of complete text analysis instead of just metadata analysis.

#### 3.4.2 Linking and Matching Citations
Clustering of documents is also performed on the basis of their citation information. For instance, papers that cite the same paper are placed in the same cluster. Combining the clustering methods, adopted using citation string parsing and versions, the cluster elements contain flag to indicate if a paper is a version or just a citation. The clustering and linking process [40] has been illustrated in Fig. 5. Every cluster has scholarly documents and citations. The arrows between clusters indicate the 'cites' relation. For instance, scholarly documents in cluster 1 cite papers in cluster 2 and cluster 3. Citation linking and matching are important step in the process in view of the fact that some fields of metadata that may have been incomplete or extracted incorrectly can be corrected and completed from the data provided by the linkage.

Considering that data will be collected from heterogeneous sources and may exist in different formats, the concept of data linking can be used. Debattista et al. [45] gave useful insights on how 'Linked Big Data' can significantly improve the veracity and value dimensions of data, also drawing parallels between methods used for big data and linked data. Linked data is a useful concept for finding events of interest and solving queries that were otherwise not possible [155][156].

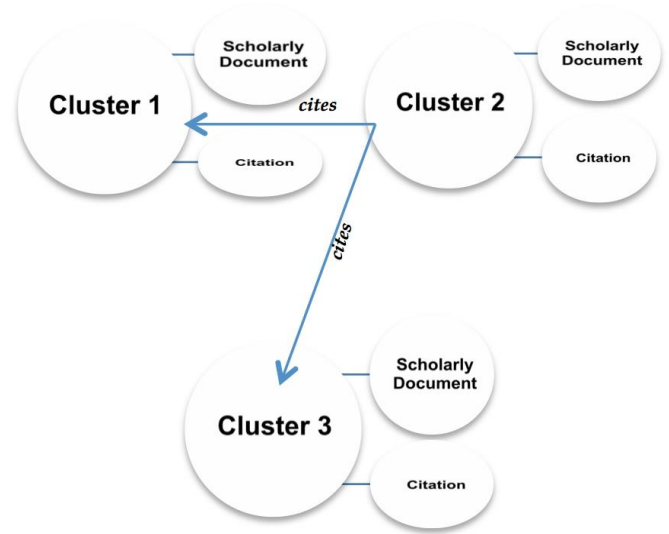

Fig. 5. Clustering and Linking of Citations and Documents

Firstly, metadata stores are populated with data, after which mapping is done using standardized resource description semantics. In order to store the metadata extracted, existing metadata formats like DBLP as Linked Data [157] may be used. The use of RDF stores seems relevant in this context. After this initial step, data can now be browsed, as it is relevant. Higher-level services can be composed to work on top of this data linking and mapping layer. Since the data will be stored in Resource Description Framework format, RDF query language (SPARQL) will have to be used for retrieving data from the data store. Hu et al. [141] makes use of this concept to drive a learning analytics web portal. Besides this, Mahmood et al. [159] gave a method for detecting document similarity, which uses RDF citation graph for social network analysis.

#### 3.4.3 Author Disambiguation
Extracted author information needs to undergo preprocessing for getting rid of the inherent ambiguity that is associated with names. Fundamentally, two issues exist in author disambiguation. Two authors may have the same name while one author may use different names. There are three types of disambiguation methods used: algorithmic, first-initial and all-initial methods.

Kim et al. [59] disambiguated the DBLP dataset using these three methods and compared their impact, concluding that author disambiguation can have a substantial influence on data quality and quality of service and analytics performed using the data. A more efficient method for disambiguation makes use of the Random Forest model and considers name, affiliation, email address and coauthors, in addition to several others [89]. Data quality and provenance management, discussed in the previous section, applies to author disambiguation as well.

### 3.5. Storage, Indexing and Processing

Once the extraction and preprocessing are complete, the extracted information needs to be ingested into the system. Distributed ingestion to address the bottleneck issues that occur during ingestion is being explored. Moreover, the data to be stored includes the original PDF along with the extracted information. Saving all this information into a single database or repository can lead to potential scalability issues. Moreover, taking backup of the single repository can take a lot of time.

A single repository built on top of an HDFS-based distributed repository can solve both these problems [111]. This will also keep the advantages of easy read and write associated with using single repository intact. Besides this, the use of graph database to store big scholarly data seems relevant owing to the linked structure of the same. There are several tools available for index maintenance, of which the most popular index maintenance engine is Solr [112]. In order to support the scalability requirements of the system, MapReduce paradigm must be used for parallelizing the extraction and ingestion processes.

### 3.6. Summary of Big Scholarly Data Management Challenges

Storing and processing an ever-increasing volume of data is a recurring challenge. Moreover, storing and processing unstructured data and performing these activities such that aggregating and correlating data from different sources become simpler, also require research attention. These challenges are inherent to any cloud-based big data analytics solution. With specific reference to big scholarly data, the challenges that persist for any system that aims to manage and process this data reserve effectively have been tabulated in Table 1.

A significant limitation that exists with respect to acquisition of data is copyright of material concerned. Khabsa and Giles [23] provided the estimation that 1 out of 4 scholarly documents are open access. It is important to note here that this is a generalized estimation and may vary from subject to subject. With that said, this issue reduces the total available data for analysis to 25%. However, this limitation shall not affect individuals and institutions that possess a copyright to access the aforementioned. A workaround framework that keeps the interest of users and copyright holders safe can be significantly valuable for researchers and scholarly community.

## 4 BIG SCHOLARLY DATA ANALYTICS

Systems need to analyze static as well as stream data. In order to create generic solutions and suffice these requirements, there is a need to integrate different programming models in the analytics engine. Moreover, energy efficiency and optimal resource usage also have to be taken into account. Specifically, there is a need for standardization in solutions and the development of most effective and efficient data processing solutions need to be emphasized [21].

Apart from scholarly documents, the big scholarly data system consists of many other types of data, which may either be generated from the extracted information or as a result of interaction between users and the system. When a user uses the system, he or she will most likely query the system. As a consequence, user statistics, querying information and logs are generated.

TABLE 1
BIG SCHOLARLY DATA MANAGEMENT CHALLENGES

| Data Management for Scholarly Resources | Challenges and Future Work |
|---|---|
| Data Acquisition | 1. Identification of sources of data as authentic and useful. 2. Differentiating between organizational and institutional sources, authors' personal webpages and other sources of data. 3. Usage and Query limits imposed by APIs limits the number of results returned. |
| Document Classification | 1. Preliminary document classification on the basis of subject or domain. |
| Data Integration | 1. Integration of heterogeneous sources of data, particularly open datasets provided by Wikipedia and Government data. |
| Information Extraction | 1. Devise methods for better accuracy, coverage and scalability. 2. Devise better methods for extraction of diverse structures. 3. Create domain-specific concept hierarchies. 4. Create a full scholarly citation graph and knowledge graph. |
| Clustering Documents and Linking Entities | 1. Devise better methods for author disambiguation. 2. Improve the quality of extracted data 3. Investigate the importance of data provenance management for big scholarly data analytics. |
| Storage, Indexing and Processing | 1. Explore the scalability of distributed processing and storage and elasticity of cloud solutions for big scholarly data. |

This data can be analyzed to get insights on user patterns, demographic analysis of system usage and system statistics. The log data maintained by the server can be mined to derive user-specific data like IP address, location of access, type of request and response returned, in addition to several others. This data can be stored in the HDFS using Hive tables and processed and queried using Pig Scripts [10].

The information extracted from the scholarly documents can be used to develop several scholarly applications. Some of the existing and well-established applications of big scholarly data analytics include research management, research paper recommendation, reviewer recommendation, collaborator discovery and expert finding. However, there is no limit to innovation. The dearth of tools and the lack of commercialization and popularity of existing tools open doors for many opportunities in this field. Existing literature on these applications have been discussed below. Table 2 gives the summary of research on scholarly applications.

## 4.1 Research Management

Research management entails a broad range of applications that are developed with the objective to facilitate research and reduce the time that researchers and scholars spend on unproductive activities by adding an element of automation in standard research guidelines and procedures. One of the best examples of a tool created for research management is RLetters [58]. This tool analyzes text inputted to it for several kinds of textual analysis like keyword co-occurrence and collocation analysis. A sample application of this tool is its use in determining if a research paper fits in the coverage of a journal, eliminating the scope of rejection caused because of such reasons.

Research is a highly dynamic activity. With research being underway all across the world in institutions, big and small, innovations happen every minute and trends change. Evidently, there is an obvious application of trends analysis and prediction in research management. Shibata et al. [78] suggested the use of topological measures for detection of new research domains in the citation network. Another aspect of research management is analysis of the impact of research, researchers and organizations.

Research is an evolution of its own kind. Therefore, the conclusions derived in one research paper may serve as inputs for future research in that area, following a linear model. However, there is a possibility that the conclusions derived in one research paper may lead to the identification of new research problems, giving rise to offshoots. Performing a correlational analysis of the topics covered by research papers can also identify research gaps and opportunities.

Some research problems exhibit transitivity. For instance, if a research paper establishes that a particular type of virus is the cause of a disease and another research paper establishes that a vaccine works for this virus, then there is a high probability that the vaccine may work for that disease. Tools can be developed for identification of research gaps that can be mathematically modeled in this manner.

TABLE 2
SUMMARY OF RESEARCH ON SCHOLARLY APPLICATIONS

| Category | Work | Goal |
|---|---|---|
| Research Management | RLetters [58] | Text analysis tool |
| | Shibata et al. [78] | Detection of new research domains |
| | Walters [117], Chen [118], Hirsch [113], Ren and Taylor [114] | Scientific impact assessment, challenges associated and applications |
| | Dong et al. [162] | Scientific impact prediction |
| | Haustein [49] | Societal impact assessment |
| Collaborator Discovery | Habib et al. [62], Kong et al. [48], Xia et al. [64], Chaiwanarom and Lursinsap [63], Yang et al. [53], Jan van Eck and Waltman [77] | Approaches for collaborator recommendations and scholars matching |
| Expert Finding | Kardan and Rafiei [152], Chen et al. [86] | Content-based approach for expert finding |
| | Widen-Wulff and Ginman [153], Widen-Wulff et al. [154] | Social Network Analysis (SNA) – based approach for expert finding |
| | Rafiei and Kardan [66], Yang et al. [65] | Hybrid approach for finding experts |
| Recommendation Systems | Research papers [121] [50] [52], Citations [51], Reviewer [141], Books [122], Academic events [131], Venues [132], News feed [129] [130], Citations for patents [133], Academic datasets [134], Educational recommendation [123] | Different types of recommendation engines |
| Other Scholarly Applications | Academic search engines [1] [2] [3] [4] [5] [6], academic alerting services [124], plagiarism detection [135] [136] [137], and research papers summarization [126][127][128] | Miscellaneous applications that make use of scholarly data |

Scholarly impact and journal reputation can be assessed using qualitative and quantitative measures, some of which are Google Scholar Metrics, Eigenfactor, Journal Citation Reports and Web of Science, in addition to several others. In order to assess the citation-based impact, several indicators like impact per publication, Source Normalized Impact per Paper, impact factor, h5-index and SCImago Journal Rank are used. Walters [117] gave a comprehensive guide to these metrics and measures. Most of the proposed methods make use of citation data for generating a ranking for organizations and scholars [113][114].

Characterization and measurement of scholarly impact suffers from several challenges in view of the fact that scholarly knowledge is a rapidly growing body. Therefore, as this data grows, it also makes some scientific contributions irrelevant, at the same time. Scientific impact prediction is another field that has attracted immense interest. Dong et al. [162] evaluated the feasibility of predicting scientific impact and proposed a model that can be used for the same purpose. However, their work is restricted to computer science and the analysis can be extended to predict which papers will be primary contributors to the predicted impact.

Chen [118] identified the challenges specific to this domain and classified them under three categories namely, creation of scientific knowledge, adaptation of the same and its diffusion. Firstly, accessibility, uncertainty and lack of standardization are the most crucial limitations. Besides this, one of the greatest challenges in the field of scholarly impact measurement is the integration of scientific metrics with analytics.

There is an increasing demand from research organizations and communities to demonstrate the societal impact of researches, much beyond their impact on the scientific community. This has led to the rise of a new term, altmetrics, which uses social media data for societal impact assessment and research evaluation. Although, this concept is still is its infancy and faces grave challenges like data quality, heterogeneity and dependencies [49], it is gradually becoming a significant part of impact analysis.

## 4.2 Collaborator Discovery

One of the popular and useful applications derived from analysis of scholarly data is collaborator discovery, which has gained all the more importance with the advent of interdisciplinary studies. There are some existing systems that support this functionality. The CiteSeerX team had implemented CollabSeer, which is a search engine that finds probable coworkers for a researcher [158]. There are several different facets of collaborator discovery that have been discussed in literature.

Firstly, collaborator discovery is a type of recommen-dation engine that matches scholars on the basis of some parameters like research interests using different approach for similarity computation to make recommendations. Out of the different approaches proposed for matching scholars, Habib et al. [62] have given one of the most recent approaches. This approach implements the inverted index using MapReduce; thus, using Universal quantifier queries on recursive relation, to match scholars. However, the implementation assumes that the inverted index created during the process fits into the main memory. In view of the fact that the dataset is considerably large, this assumption may not be true, which fuels the need to explore ways in which this intermediate data can be distributed and managed.

Many factors like publication contents and collaboration networks [48] and academic factors like coauthor order and collaboration parameters [64] have been exploited for modeling the problem. In view of the fact that this application finds its roots in interdisciplinary nature of research problems, Chaiwanarom and Lursinsap [63] used degrees of collaborative forces, seniority and evolution of research interest for recommendation. While most of the previous researches in this area concentrate on social proximity analysis, Yang et al. [53] proposed an approach for making recommendations in heterogeneous bibliographic networks by considering not just social proximity, but also institutional connectivity, adding a degree of intelligence to the process.

Jan van Eck and Waltman [77] undertook an extensive review on spatial scientometric data analysis and concluded that most studies present a national level analysis, not detailing it to the regional and urban levels. Such an analysis can be crucial for collaborations in which location of the collaborators are crucial. Therefore, future studies can incorporate this facet of collaborator discovery and recommendation.

## 4.3 Expert Finding

Finding experts is a concept that was mostly focused upon by organization. However, lately, there has been an increasing shift in research interests towards finding experts in online communities and social networks [125][150]. Formally named as Expert Finder Systems (EFSs), these systems form a specialized class of recommender systems [151]. There are two basic approaches followed for implementation of these systems namely, content-based approach and Social Network Analysis (SNA) – based approach. While the former makes use of text mining techniques [152][86], the latter focuses on concepts like PageRank and HITS for identifying experts [153][154].

Rafiei and Kardan [66] make use of a hybrid approach, using content analysis (Concept Map) as well as social network analysis (PageRank) for finding experts. The use of semantic network based methods for computing similarity results in high precision and good results. Most of the existing systems mine individual-level information for identifying experts. However, many other measures can be used to extract semantic similarity for improved accuracy. In order to broaden the scope and coverage and improving the specificity of results, Yang et al. [65] scans scholars for information about social network of the individual, research relevance and institutional connectivity for recommending an expert.

### 4.4 Other Recommender Systems

The concept of recommendation systems finds important applications in the field of big scholarly data. Several types of recommender systems can be used to recommend research papers, books [122], academic events [131], venues [132], news feed [129] [130], citations for patents [133] and academic datasets [134]. Brusilovsky et al. [123] have also introduced the concept of educational recommendation. In addition to this, some applications like academic search engines, academic alerting services [124], plagiarism detection [135] [136] [137], and research papers summarization [126][127][128] also exist.

From the first research paper recommendation system introduced by Bollacker et al. [121] in the year 1998, there have been many proposed and implemented systems in this area. Beel et al. [61] gave an extensive review on the work performed on research paper recommendation systems. The main findings of this survey were that most of the systems were mere proposals for which no implementation even came into existence. As a result, it is difficult to make any comparisons. This led to the realization of the need for an evaluation system. Besides this, most of the implemented systems used accuracy as the testing parameter, which is rather incomplete in view of the fact that user experience and usability are equally important parameters.

Ismail and Al-Feel [50] proposed a Hadoop-based recommendation system for research papers, which is specifically designed for digital libraries. A comparatively lesser-explored area is the integration of mind mapping tools with recommendation systems. Beel et al. [52] explored this possibility by proposing an approach that models users on the basis of mind maps and evaluated their approach using Docear, a reference management system.

Closely related to the discussion is RefSeer [51], a citation recommendation system that supports global and local recommendation. For global recommendation, a topic modeling-based topical composition is computed from the text [119]. On the other hand, the citation translational model is used for making local recommendations [120]. West et al. [60] introduced the concept of Eigenfactor Recommends, a citation-based method for improving scholarly navigation. The algorithm uses the hierarchical structure of scientific knowledge, making possible multiple scales of relevance for different users. The approach presented in this paper shares resemblance to the co-citation approach. However, the coverage achieved by the former is better than that of the latter.

Most academic search engines provide research paper recommendation as an additional service to their users. Academic engines and paper recommendation systems are essentially based on same methodology and uses the same set of techniques [227][228][229]. The idea is to calculate the similarity between user queries and documents. On the other hand, academic engines compute research interests and then calculate the similarity between available documents and computed research interests to make recommendations. Reviewer recommendation systems are based on the fact that the scholars who have research papers in specific areas can be considered reviewers for other papers belonging to the same area [141].

The only difference between research paper recommendation and reviewer recommendation is that the former scans a corpus of papers to suggest papers that match research interests of the concerned scholar while the latter scans scholars to give a list of scholars who have published in the same area as the research paper to be reviewed. Wang et al. [142] presented an extensive review on the reviewer assignment problem.

Scientometrics, a field that deals with the study of scholarly impact also finds relevance in the research paper recommendation systems context. Several metrics like h-index [143], bibliographic coupling strength [144] and co-citation strength [145] have found applications in recommendation systems [146][147][148][149]. Besides these, collaborative [140] and content-based [138][139] filtering from other domains like news and movies is also used in recommendation systems.

## 5 VISUALIZATION

Broadly, in the area of visualization and user interaction, real-time visualization of data is an important area of research. The research community is yet to devise solutions that can visualize data at the rate at which the same is generated and in the amounts that it exists. Parallel research in the development of cost-effective devices for large-scale visualization is also underway [21].

With specific reference to scholarly data, visualization poses several challenges. Visualization for scholarly applications can be viewed as a subset of visualization for learning analytics for the sheer similarity that these two fields share in their objective. Apart from many others, one of the most significant factors that must be paid heed to is visualization of uncertainty. Uncertainty is an invincible aspect and result of every phase of the system.

Moreover, uncertainty, when visualized appropriately and effectively can be a great aid for decision-making. Demmans Epp and Bull [33] provided a survey that indicated the importance of representing uncertainty in learning analytics applications and suggested ways in which existing visualizations can be augmented for the same. The viability of this concept for scholarly applications is suggested as future research in this area.

An effective visualization is fundamental to any scholarly application. One such application, designed by Widyantoro and Oenang [34], enabled the user to visualize his or her research map. Although, this is a very basic system, it can be improved and integrated with a research management system to make it easy for scholars to manage and perform research.

Another area of research specific to scholarly data is visualization of bibliometric networks. Citation, co-authorship, co-citation, keyword co-occurrence and bibli-ographic coupling, in addition to several other types of networks concerning scholarly data are termed as bibliographic networks [77]. Nakazawa et al. [76] proposed a topic-based clustering technique for visualization of citation networks. Kiado et al. [75] performed preliminary research in this field and proposed a method for identification and visualization of research groups on the basis of factorial analysis of raw data and similarity in choice of co-authors.

Khalid et al. [74] explored the generation of large dynamic networks, which is a requirement of citation network. The proposed method makes use of Pajek tool that has been extended to create a set of JUNG libraries. Co-authorship network is the other type of network that needs to be created using scholarly dataset. Tools used for bibliometric network analysis have been explained in Table 3.

TABLE 3
VISUALIZATION TOOLS FOR BIBLIOMETRIC NETWORK ANALYSIS

| Tool | Features |
|------|----------|
| Pajek [85] | General-purpose network analysis tool for visualization of large networks. |
| Gephi [79] | General-purpose network analysis tool for visualization of dynamic networks and complex systems. |
| VOSviewer [80] | It is a software tool that supports text mining. Therefore, it is used for visualization of co-occurrence networks. |
| HistCite [81] | It is a Windows-based software package for information visualization and bibliometric analysis. |
| CiteSpace [82] | It is a Java application used for analysis of patterns and trends in scientific literature. |
| CitNetExplorer [83] | It is a software tool used for citation network analysis. |
| Sci2 [84] | It is a modular toolset that supports visualization and topical, network, geospatial and temporal analysis of scholarly datasets at global, local and micro levels. |

# 6 OTHER OPEN CHALLENGES

The non-technical challenges are further classified into business-related challenges and miscellaneous challenges. The former category of challenges includes the need to make these solutions cost-effective and the inability of the available solutions to replicate analyses and create generic solutions. Besides this, the lack of staff and debugging and testing solutions are some of the miscellaneous challenges faced.

Most organizations and institutions have existing digital libraries. This can serve as a solution to the copyright issues as these organizations have licenses to access copyrighted content. Therefore, analytical services and applications can be provisioned as products that can incorporate existing digital libraries of the institute and integrate it with the huge Internet data reserve to serve Intranet users for increased usability and commercial viability. However, this shall need development of APIs and solutions that can support this kind of functionality.

In addition to this, the lack of scholars' engagement in social platforms is a limitation and challenge for designing next-generation platforms for collaborations. However, with the increasing popularity of social scholarly platforms like ResearchGate, things are rapidly changing. Veletsianos and Kimmons [115] have presented an analysis of scholars' engagement and usage patterns on Twitter. The relevance of such studies in measuring scholarly impact needs to be explored. This opens doors for many scholarly applications and their usability in the existing scenario.

# 7 CONCLUSION

This survey includes a detailed study of the current trends and existing challenges in the different subsystems of the big scholarly data platform, with specific focus on directions for future research in this area. The challenges have been divided into two fundamental categories namely, technical and non-technical challenges. Since, the paper focuses on technical challenges, this category has been further divided into three categories namely, data management, analytics and visualization. All these categories have been individually covered in different sections of the paper.

Several studies suggest that cloud computing is an apt solution for the big data problem. However, there are several issues that need to be addressed before this synergistic model can be called commercially viable. Suggested future work in the area includes the development of solutions and APIs. Moreover, the user must be able to switch among the available solutions. Secondly, the real potential of cloud computing and the elasticity that it offers is yet to be explored. Most of the future work in this direction includes creation of expressive languages that shall enable users to define their problem to the system keeping in view that operational efficiency of the system with the increasing data only needs to get better.

Scholarly data is a huge data reserve, which is substantially appended on a daily basis and includes a variety of data. As a result, it is popularly termed as big schol-arly data. Several applications can be designed using analysis and visualization of this data. With specific reference to big scholarly data platform, challenges and limitations exist at every stage of the data analytics process. Research is underway in specific components of this platform, which needs to be integrated for the development of a comprehensive system.

While CiteSeerX exists as one of the most popular scholarly platforms, the services provided are rather limited in their functionality and can be further enhanced to include many scholarly applications like research management and optimized to provide added functionality like algorithm linking, time-evolution of research and recommendations. Moreover, there is a lack of tools and techniques that can facilitate research and automate unproductive aspects involved in the process, paving way for innovation.

## ACKNOWLEDGMENT

This work was supported by a grant from "Young Faculty Research Fellowship" under Visvesvaraya PhD Scheme for Electronics and IT, Department of Electronics & Information Technology (DeitY), Ministry of Communications & IT, Government of India.